\documentclass[conference,a4paper]{IEEEtran}

\usepackage{amsfonts}
\usepackage{enumerate}
\usepackage{cite}
\usepackage{times}
\usepackage{amssymb}
\usepackage{mathrsfs}
\usepackage{amsmath}
\usepackage{cite}
\usepackage[ruled,vlined]{algorithm2e}
\usepackage{tikz}
\usetikzlibrary{arrows,automata}
\usepackage[latin1]{inputenc}
\usepackage{verbatim}
\makeatletter

\newcommand{\Rmnum}[1]{\expandafter\@slowromancap\romannumeral #1@}

\makeatother

\newtheorem{thm}{Theorem}
\newtheorem{lemma}[thm]{Lemma}

\newtheorem{cor}[thm]{Corollary}
\newtheorem{defn}{Definition}
\newtheorem{rem}[thm]{Remark}

\newtheorem{rem-eg}[thm]{Remark and Example}

\newcommand{\mF}{\mathcal{F}}
\newcommand{\mP}{\mathcal{P}}

\newcommand{\Rank}{{\mathrm{Rank}}}

\SetKw{KwInitialization}{Initialization:}

\begin{document}

\sloppy

\title{The Failure Probability of Random Linear Network Coding for Networks}

\author{
  \IEEEauthorblockN{Xuan~Guang}
  \IEEEauthorblockA{School of Mathematical Science and LPMC\\
    Nankai University\\
    Tianjin, P. R. China\\
    Email: xguang@nankai.edu.cn}
  \and
  \IEEEauthorblockN{Fang-Wei~Fu}
  \IEEEauthorblockA{Chern Institute of Mathematics and LPMC\\
    Nankai University\\
    Tianjin, P. R. China\\
    Email: fwfu@nankai.edu.cn}
}



\maketitle

\begin{abstract}
In practice, since many communication networks are huge in scale, or complicated in structure, or even dynamic, the predesigned linear network codes based on the network topology is impossible even if the topological structure is known. Therefore, random linear network coding has been proposed as an acceptable coding technique for the case that the network topology cannot be utilized completely. Motivated by the fact that different network topological information can be obtained for different practical applications, we study the performance analysis of random linear network coding by analyzing some failure probabilities depending on these different topological information of networks. We obtain some tight or asymptotically tight upper bounds on these failure probabilities and indicate the worst cases for these bounds, i.e., the networks meeting the upper bounds with equality. In addition, if the more topological information of the network is utilized, the better upper bounds are obtained. On the other hand, we also discuss the lower bounds on the failure probabilities.
\end{abstract}

\section{Introduction}

Network coding was proposed by Ahlswede \emph{et al.} \cite{Ahlswede-Cai-Li-Yeung-2000}, which shows that if coding is applied at the nodes instead of routing alone, the source node can multicast the information to all sink nodes at the theoretically maximum rate. Li \emph{et al.} \cite{Li-Yeung-Cai-2003} further indicated that linear network coding with finite alphabet size is sufficient for multicast. Koetter and M\'{e}dard \cite{Koetter-Medard-algebraic} presented an algebraic characterization of network coding. Although network coding allows the higher information rate than classical routing, Jaggi \emph{et al.} \cite{co-construction} still proposed a deterministic polynomial-time algorithm for constructing a linear network code.
Random linear network coding was introduced by Ho \emph{et al.} \cite{Ho-etc-random} as an acceptable coding technique for many communication problems, particularly, for the case that the network topology cannot be utilized completely, because it is impossible to use predesigned network codes. Their main results are upper bounds on different failure probabilities which characterize the performance of random linear network coding. Balli \emph{et al.} \cite{zhang-random} improved on these bounds and analyzed their asymptotic behavior as the field size goes to infinity. However, these upper bounds are not tight. In order to characterize the performance of random linear network coding more comprehensively and completely, Guang and Fu \cite{Guang-aver-ptob-IEICE} introduced and studied the average failure probability of random linear network coding. In this paper, we further discuss the random linear network coding and improve on the bounds for different cases. In particular, if the more knowledge about the topology of the network is known, we can obtain the better bounds. Further, we indicate that these bounds are either tight or asymptotically tight.

A communication network is represented by a finite acyclic directed graph $G=(V,E)$, where $V$ and $E$ are the sets of nodes and channels of the network, respectively. A direct edge $e=(i,j)\in E$ stands for a channel leading from node $i$ to node $j$. Node $i$ is called the tail of $e$ and node $j$ is called the head of $e$, denoted by $tail(e)$ and $head(e)$, respectively. Correspondingly, the channel $e$ is called an outgoing channel of $i$ and an incoming channel of $j$. For a node $i$, define $Out(i)=\{e\in E:\ tail(e)=i\}$, and $In(i)=\{e \in E:\ head(e)=i\}$.
We allow the multiple channels between two nodes and assume that one field symbol can be transmitted over a channel in a unit time. In this paper, we only consider networks with single source, and the unique source node is denoted by $s$, which generates messages and transmits them to all sink nodes $t\in T$ by network coding, where $T$ is the set of sink nodes. Denote $C_t$ the minimum cut capacity between the source node $s$ and the sink node $t$. Let the information rate be $w$ symbols per unit time which means that the source messages are $w$ symbols $\underline{\bf{X}}=(X_1,X_2,\cdots,X_w)$ arranged in a row vector where each $X_i$ is an element of the finite base field $\mathcal{F}$. In this paper, we always assume that $w\leq C_t$ for any $t\in T$. We use $U_e$ to denote the message transmitted over channel $e=(i,j)$ and
$U_e$ is calculated by the following formula $U_e=\sum_{d\in In(i)}k_{d,e}U_d$, where $k_{d,e}\in \mF$ is called the local encoding coefficient for the adjacent pair of channels $(d,e)$. Further, it is not difficult to see that $U_e$, actually, is a linear combination of the $w$ source symbols $X_i$, $1\leq i\leq w$, that is, there is an $w$-dimensional column vector $f_e$ over the base field $\mF$ such that $U_e=\underline{\bf{X}}\cdot f_e$ (see \cite{Zhang-book}\cite{Yeung-book}). This column vector $f_e$ is called the global encoding kernel of a channel $e$ and it can be determined by the local encoding coefficients. Further, at the sink node $t\in T$, all global encoding kernels and received messages of incoming channels are available. Define an $w\times |In(t)|$ matrix $F_t$ and an $|In(t)|$-dimensional vector $A_t$ as $F_t=(f_e:\ e\in In(t))$ and $A_t=(U_e:\ e\in In(t))$. Then we have decoding equation $A_t=\underline{\bf{X}}\cdot F_t$, which implies that the sink node $t$ can decode (recover) the original source message vector $\underline{\bf{X}}$ successfully if and only if $\Rank(F_t)=w$.

The main idea of random linear network coding is that when a node (maybe the source node $s$) receives the messages from its all incoming channels, for each outgoing channel, it randomly and uniformly picks the encoding coefficients from the base field $\mF$, uses them to encode the received messages, and transmits the encoded messages over the outgoing channel. In other words, the local coding coefficients $k_{d,e}$ are independently and uniformly distributed random variables taking values in the base field $\mF$. Since random linear network coding does not consider the global network topology or coordinate coding at different nodes, it may not achieve the best possible performance of network coding, that is, some sink nodes may not decode correctly. Therefore, the performance analysis is very important both theoretically and for application. Before proceeding further, we first introduce the definitions of the failure probabilities in order to characterize the performance analysis of random linear network coding.
\begin{defn}
For random linear network coding on $G$,
\begin{itemize}
   \item $P_e\triangleq Pr(\exists\ t\in T \mbox{ such that }\Rank(F_t)<w)$ is called
the failure probability of random linear network coding for network $G$, that is the probability that
the messages cannot be decoded correctly at at least one sink node
in $T$.
    \item $P_{e_t}\triangleq Pr(\Rank(F_t)<w)$ is called the failure
probability of random linear network coding at sink node $t$, that is the probability that the
source messages cannot be decoded correctly at the sink node $t\in T$,
\end{itemize}
\end{defn}


\section{Failure Probability for Networks}
In this section, we will present our main results on the failure probability of random linear network coding for network $G$, where $G$ is any fixed network with single source $s$. Let $T=\{t_1,t_2,\cdots,t_l\}$ be the set of sink nodes. For each sink node $t_i\in T$, by Menger's Theorem, there exist $w$ channel disjoint paths from $s$ to $t_i$ as $w\leq C_{t_i}$. Denote the collection of the arbitrarily chosen $w$ paths for $t_i$ by $\mP_i=\{P_{i,1},P_{i,2},\dots,P_{i,w}\}$, where the path $P_{i,j}=\{e_{i,j,1},e_{i,j,2},\cdots,e_{i,j,m_{i,j}}\}$ satisfying $tail(e_{i,j,1})=s$, $head(e_{i,j,m_{i,j}})=t_i$, and $tail(e_{i,j,k})=head(e_{i,j,k-1})$ for others. Obviously, it is possible that $\mP_i \cap \mP_j \neq \emptyset$ for distinct sink nodes $t_i$ and $t_j$.
Let $r_i$ be the number of the internal nodes in $\mP_i$ and $R$ be the number of the internal nodes in $\cup_{t_i\in T}\mP_i=\cup_{i=1}^{l}\mP_i$. Clearly, $\max_{1\leq i \leq l}r_i\leq R \leq \sum_{i=1}^{l}r_i$. Denote the $R$ internal nodes by $i_1,i_2,\cdots,i_R$ and let the ancestrally topological order be
$s\triangleq i_0 \prec i_1 \prec i_2 \prec \cdots \prec i_R \prec \{t_1,t_2,\cdots,t_l\}.$

During our discussion, we use the concept of cuts of the paths from $s$ to $t$ introduced in \cite{zhang-random} and \cite{Guang-aver-ptob-IEICE}, which is different from the concept of cuts of networks in graph theory. For each $t_i$, the first cut $CUT_{i,0}$ is the set of the $w$ imaginary channels, i.e., $CUT_{i,0}=In(s)$. At $s$, the next cut $CUT_{i,1}$ is the set of the first channels of all $w$ paths, i.e., $CUT_{i,1}=\{e_{i,1,1},e_{i,2,1},\cdots,e_{i,w,1}\}$. At node $i_1$, the next cut $CUT_{i,2}$ is formed from $CUT_{i,1}$ according to the following method: if $In(i_1)\cap CUT_{i,1}\neq \emptyset$, then replace the channels in $In(i_1)\cap CUT_{i,1}$ by their respective next channels in the paths, other channels remain the same as in $CUT_{i,1}$; otherwise if $In(i_1)\cap CUT_{i,1}=\emptyset$, $CUT_{i,2}$ remains the same as $CUT_{i,1}$. In the same way, once $CUT_{i,k}$ is defined, $CUT_{i,k+1}$ is formed from $CUT_{i,k}$ by the same method above. By induction, all cuts $CUT_{i,k}$ can be defined for $i=1,2,\cdots,l$ and $k=0,1,\cdots,R+1$. Particularly, note that $CUT_{i,R+1}=\{e_{i,1,m_{i,1}},e_{i,2,m_{i,2}}, \cdots, e_{i,w,m_{i,w}}\}$, that is the set of the last channels of all $w$ paths from $s$ to $t_i$. Further, for each node $i_k$, $k=0,1,2,\cdots,R$, define $CUT_{t,k}^{out}=\{e:\ e\in CUT_{t,k}\setminus In(i_{k})\}$, and two sets $M_{k}=\{ t_i: CUT_{i,k}\neq CUT_{i,k+1} \}$ and $N_k=\{ t_i: CUT_{i,k}\neq CUT_{i,k+1}, \mbox{ and } CUT_{i,k+1}=CUT_{i,R+1} \}$. In fact, $M_{k}$ is the set of sink nodes satisfying that at least one of its $w$ paths passes through the node $i_k$, and $N_k$ is the set of the sink nodes satisfying that at least one of its $w$ paths passes through the node $i_k$ and $i_k$ is the last internal node on its $w$ paths. Furthermore, let $|M_k|=m_k$ and $|N_k|=n_k$. Then $\sum_{k=0}^{R}n_k=l$, $\sum_{k=0}^{R}m_k=(r_1+1)+(r_2+1)+\cdots+(r_l+1)=\sum_{i=1}^{l}r_i+l$, and thus $\sum_{k=0}^{R}(m_k-n_k)=\sum_{i=1}^{l}r_i$.

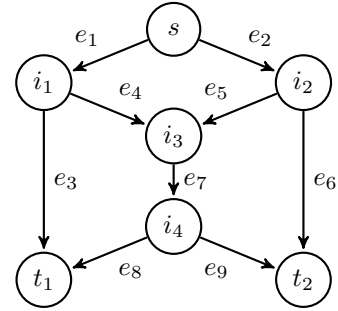
\begin{figure}[!h]
\begin{center}
\begin{tikzpicture}[->,>=stealth',shorten >=1pt,auto,node distance=1cm,
                    thick]
  \tikzstyle{every state}=[fill=none,draw=black,text=black,minimum size=7mm]
  \node[state]         (s)                    {$s$};
  \node[state]         (i1) [below left of=s,xshift=-10mm] {$i_1$};
  \node[state]         (i2) [below right of=s,xshift=10mm] {$i_2$};
  \node[state]         (i3) [below right of=i1,xshift=10mm] {$i_3$};
  \node[state]         (i4) [below of=i3,yshift=-2mm]       {$i_4$};
  \node[state]         (t1) [below left of=i4,xshift=-10mm]   {$t_1$};
  \node[state]         (t2) [below right of=i4,xshift=10mm]  {$t_2$};
  \path (s) edge              node [swap]{$e_1$} (i1)
            edge              node {$e_2$} (i2)
        (i1)edge              node {$e_3$} (t1)
            edge              node {$e_4$} (i3)
        (i2)edge              node[swap] {$e_5$} (i3)
            edge              node {$e_6$} (t2)
        (i3)edge              node {$e_7$} (i4)
        (i4)edge              node {$e_8$} (t1)
        (i4)edge              node[swap] {$e_9$} (t2);
\end{tikzpicture}
\caption{Butterfly Network $G_1$}
\label{fig_bn_2}
\end{center}
\end{figure}

In order to illustrate the concepts introduced, we take the butterfly network $G_1$ (Fig.\ref{fig_bn_2}) as an example.
For sink nodes $t_1, t_2\in T$, let $\mP_{1}=\{P_{1,1},P_{1,2}\}$ and $\mP_{2}=\{P_{2,1},P_{2,2}\}$, where $P_{1,1}=\{ e_1,e_3 \}$, $P_{1,2}=\{ e_2,e_5,e_7,e_8 \}$, $P_{2,1}=\{ e_1,e_4,e_7,e_9 \}$, and $P_{1,2}=\{ e_2,e_6 \}$. Then
\begin{align*}
&CUT_{1,0}=\{d_1,d_2\},\ CUT_{1,0}^{out}=\emptyset,\\
&CUT_{1,1}=\{e_1,e_2\},\ CUT_{1,1}^{out}=\{e_2\},\\
&CUT_{1,2}=\{e_3,e_2\},\ CUT_{1,2}^{out}=\{e_3\},\\
&CUT_{1,3}=\{e_3,e_5\},\ CUT_{1,3}^{out}=\{e_3\},\\
&CUT_{1,4}=\{e_3,e_7\},\ CUT_{1,4}^{out}=\{e_3\},\\
&CUT_{1,5}=\{e_3,e_8\},\ CUT_{1,5}^{out}=\emptyset;\\
&CUT_{2,0}=\{d_1,d_2\},\ CUT_{2,0}^{out}=\emptyset,\\
&CUT_{2,1}=\{e_1,e_2\},\ CUT_{2,1}^{out}=\{e_2\},\\
&CUT_{2,2}=\{e_4,e_2\},\ CUT_{2,2}^{out}=\{e_4\},\\
&CUT_{2,3}=\{e_4,e_6\},\ CUT_{2,3}^{out}=\{e_6\},\\
&CUT_{2,4}=\{e_7,e_6\},\ CUT_{2,4}^{out}=\{e_6\},\\
&CUT_{2,5}=\{e_9,e_6\},\ CUT_{2,5}^{out}=\emptyset;\\
&M_0=M_1=M_2=M_3=M_4=\{t_1,t_2\},\\
&N_0=N_1=N_2=N_3=\emptyset,\ N_4=\{t_1,t_2\}.
\end{align*}
\begin{thm}\label{thm_cohe_net}
The failure probability of random linear network coding for the network $G$ satisfies:
$$P_e\leq 1-(1-a)^{l}\prod_{k=0}^{R-1}[1-(m_k-n_k)a],$$
where $a\triangleq 1-\prod_{h=1}^w(1-\frac{1}{|\mathcal{F}|^h})$.
\end{thm}

Before giving the proof, we need the following lemma.
\begin{lemma}{\label{lem_bound}}
Let $\mathcal{L}$ be an $n$-dimensional linear space over a finite field $\mathcal{F}$,
$\mathcal{L}_0,\ \mathcal{L}_1$ be two subspaces of $\mathcal{L}$
of dimensions $k_0,\ k_1$, respectively, and
$\langle\mathcal{L}_0\cup\mathcal{L}_1\rangle=\mathcal{L}$. Let
$l_1,\ l_2,\ \cdots,\ l_{n-k_0}$ be $(n-k_0)$ independently and
uniformly distributed random vectors taking values in
$\mathcal{L}_1$. Then
$$Pr(\dim(\langle \mathcal{L}_0 \cup \{l_1,\ \cdots,\ l_{n-k_0}\}\rangle)=n)=\prod_{i=1}^{n-k_0}\left(
1-\frac{1}{\mathcal{|F|}^i}\right).$$
\end{lemma}
\begin{IEEEproof}[Proof of Theorem \ref{thm_cohe_net}]
For each sink node $t_i\in T$, recall that the matrix $F_{t_i}=(f_e:e\in In(t))$ of size $w\times |In(t)|$ is the decoding matrix of $t_{i}$, which further is denoted by $F_i$ for convenience. Further, Define an $w\times w$ matrix $F_i'=(f_{e_{i,1,m_{i,1}}},f_{e_{i,2,m_{i,2}}},\cdots,f_{e_{i,w,m_{i,w}}})$, where recall that $e_{i,j,m_{i,j}}$, $1\leq j \leq w$, are the last channels of the chosen $w$ channel disjoint paths from $s$ to $t_i$. It is readily seen that $F_i'$ is a submatrix of $F_i$. So the event ``$\Rank(F_i)<w$'' implies the event ``$\Rank(F_i')<w$'', Hence
$$
P_e= Pr(\cup_{i=1}^{l}\Rank(F_i)<w)\leq Pr(\cup_{i=1}^{l}\Rank(F_i')<w).
$$
In addition, let $F_i^{(k)}=(f_e: e\in CUT_{i,k})$ be $w\times w$ matrices for $i=1,2,\cdots,l$ and $k=0,1,\cdots, R+1$. If $\Rank(F_i^{(k)})<w$, we call that we have a failure at $CUT_{i,k}$ and the event ``$\Rank(F_i^{(k)})=w$'' is denoted by $\Gamma_{i,k}$. Note that $F_i'=F_i^{(R+1)}$ since $CUT_{i,R+1}=\{e_{i,1,m_{i,1}},e_{i,2,m_{i,2}},\cdots,e_{i,w,m_{i,w}}\}$.
Then it further follows that
\begin{align*}
P_e \leq
1-Pr(\cap_{i=1}^{l}\Rank(F_i')=w)=1-Pr(\cap_{i=1}^{l}\Gamma_{i,R+1}).
\end{align*}
Next, we consider the probability $Pr(\cap_{i=1}^{l}\Gamma_{i,R+1})$. First,
\begin{align}
&Pr(\cap_{i=1}^{l}\Gamma_{i,R+1})\nonumber\\
\geq& Pr(\cap_{i=1}^{l}\Gamma_{i,R+1},\cap_{i=1}^{l}\Gamma_{i,R},\cdots,
\cap_{i=1}^{l}\Gamma_{i,1},\cap_{i=1}^{l}\Gamma_{i,0})\nonumber\\
=&Pr(\cap_{i=1}^{l}\Gamma_{i,0})\cdot \prod_{k=0}^{R}Pr(\cap_{i=1}^{l}\Gamma_{i,k+1}|\cap_{i=1}^{l}\Gamma_{i,k})\label{5}\\
=&\prod_{k=0}^{R}Pr(\cap_{i=1}^{l}\Gamma_{i,k+1}|\cap_{i=1}^{l}\Gamma_{i,k}),\label{2_1}
\end{align}
where (\ref{5}) follows because encoding at any node is independent of what happened before this node
as long as no failure has occurred up to this node, and (\ref{2_1}) follows from
$Pr(\cap_{i=1}^{l}\Gamma_{i,0})=Pr(\Rank(I_{w\times w})=w)\equiv1$.
Subsequently, we take into account the probability $Pr(\cap_{i=1}^{l}\Gamma_{i,k+1}|\cap_{i=1}^{l}\Gamma_{i,k})$. Actually,
\begin{align}
&Pr(\cap_{i=1}^{l}\Gamma_{i,k+1}|\cap_{i=1}^{l}\Gamma_{i,k})=Pr(\cap_{t_i\in M_k}\Gamma_{i,k+1}|\cap_{i=1}^{l}\Gamma_{i,k})\nonumber\\
=&Pr(\cap_{t_j\in M_k-N_k}\Gamma_{j,k+1}|\cap_{i=1}^{l}\Gamma_{i,k})\nonumber\\
&\cdot \prod_{t_i\in N_k}Pr(\Gamma_{i,k+1}|\cap_{i=1}^{l}\Gamma_{i,k})\label{2_2}\\
=&Pr(\cap_{t_j\in M_k-N_k}\Gamma_{j,k+1}|\cap_{i=1}^{l}\Gamma_{i,k})\prod_{t_i\in  N_k}Pr(\Gamma_{i,k+1}|\Gamma_{i,k}),\nonumber
\end{align}
where (\ref{2_2}) follows because for $t_i\in N_k$, the event $\Gamma_{i,k+1}$ is conditional independent with $\cap_{t_j\in M_k-\{t_i\}}\Gamma_{j,k+1}$ under the condition $\cap_{i=1}^{l}\Gamma_{i,k}$. Reasonably, put $\prod_{t_i\in  N_k}Pr(\Gamma_{i,k+1}|\Gamma_{i,k})=1$ for $N_k=\emptyset$, and put
$Pr(\cap_{t_j\in M_k-N_k}\Gamma_{j,k+1}|\cap_{i=1}^{l}\Gamma_{i,k})=1$ for $M_k-N_k=\emptyset$.
Further applying Lemma \ref{lem_bound}, one has
\begin{align*}
&Pr(\cap_{t_j\in M_k-N_k}\Gamma_{j,k+1}|\cap_{i=1}^{l}\Gamma_{i,k})\\
=&1-Pr(\cup_{t_j\in M_k-N_k}\Gamma_{j,k+1}^{c}|\cap_{i=1}^{l}\Gamma_{i,k})\\
\geq& 1-\sum_{t_j\in M_k-N_k}Pr(\Gamma_{j,k+1}^{c}|\cap_{i=1}^{l}\Gamma_{i,k})\\
=&1-\sum_{t_j\in M_k-N_k}[1-Pr(\Gamma_{j,k+1}|\cap_{i=1}^{l}\Gamma_{i,k})]\\
=&1-\sum_{t_j\in M_k-N_k}[1-Pr(\Gamma_{j,k+1}|\Gamma_{j,k})]\\
=&1-\sum_{t_j\in M_k-N_k}\left[1-\prod_{h=1}^{w-|CUT_{j,k}^{out}|}\left(1-\frac{1}{|\mF|^h}\right)\right]\\
\geq&1-(m_k-n_k)\left[1-\prod_{h=1}^{w}\left(1-\frac{1}{|\mF|^h}\right)\right]=1-(m_k-n_k)a.
\end{align*}
On the other hand,
\begin{align*}
&\prod_{t_i\in  N_k}Pr(\Gamma_{i,k+1}|\Gamma_{i,k})=\prod_{t_i\in  N_k}\prod_{h=1}^{w-|CUT_{i,k}^{out}|}\left(1-\frac{1}{|\mF|^h}\right)\\
\geq&\prod_{t_i\in  N_k}\prod_{h=1}^{w}\left(1-\frac{1}{|\mF|^h}\right)
=\left[\prod_{h=1}^{w}\left(1-\frac{1}{|\mF|^h}\right)\right]^{n_k}\\
=&(1-a)^{n_k}.
\end{align*}
Combining the above, it follows that
$$Pr(\cap_{i=1}^{l}\Gamma_{i,k+1}|\cap_{i=1}^{l}\Gamma_{i,k})\geq (1-a)^{n_k}(1-(m_k-n_k)a),$$
which implies that
\begin{align}
&Pr(\cap_{i=1}^{l}\Gamma_{i,R+1})\geq\prod_{k=0}^{R}Pr(\cap_{i=1}^{l}\Gamma_{i,k+1}|\cap_{i=1}^{l}\Gamma_{i,k})
\nonumber\\
\geq&\prod_{k=0}^{R}(1-a)^{n_k}(1-(m_k-n_k)a)\nonumber\\
=&(1-a)^{\sum_{k=0}^{R} n_k}\prod_{k=0}^{R}(1-(m_k-n_k)a)\nonumber\\
=&(1-a)^{l}\prod_{k=0}^{R-1}(1-(m_k-n_k)a),\label{6}
\end{align}
where the last equality (\ref{6}) follows from $\sum_{k=0}^{R}n_k=l$ and $m_{R}=n_{R}$. So the proof is completed.
\end{IEEEproof}

\begin{rem}
This upper bound on the failure probability for network is achievable. We will give a specific network below to show the tightness. For a given information rate $w$, the network $G_2$ is constructed as follows. Let the unique source node be $s$,
the sink nodes be $t_1,t_2,\cdots,t_l$. Construct a plait network $G_1'$ (see Fig. \ref{fig_plait_net}) with $R$ internal nodes for the sink node $t_1$, and plait networks $G_j'$ without internal nodes for other sink nodes $t_j\ (j=2,3,\cdots,l)$. These $l$ plait networks share a common source node $s$, i.e., the network $G_2$ is the union of the $l$ plait networks $G_j'\ (j=1,2,3,\cdots,l)$.
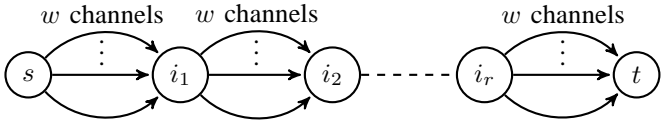
\begin{figure}[!htb]
\begin{center}
\begin{tikzpicture}[->,>=stealth',shorten >=1pt,auto,node distance=2cm,
                    thick]
  \tikzstyle{every state}=[fill=none,draw=black,text=black,minimum size=6mm]
  \tikzstyle{place}=[fill=none,draw=white,minimum size=0.1mm]
  \node[state]         (s)                 {$s$};
  \node[state]         (i_1)[right of=s]   {$i_1$};
  \node[state]         (i_2)[right of=i_1]   {$i_2$};
  \node[state]         (i_r)[right of=i_2]   {$i_r$};
  \node[state]         (t)[right of=i_r]     {$t$};
  \path (s) edge [bend left=45]          node {$w$ channels} (i_1)
            edge                         node {$\vdots$} (i_1)
            edge [bend right=45]         node {} (i_1)
      (i_1) edge [bend left=45]          node {$w$ channels} (i_2)
            edge                         node {$\vdots$} (i_2)
            edge [bend right=45]         node {} (i_2)
      (i_2) edge [-,dashed]                node {} (i_r)
      (i_r) edge [bend left=45]          node {$w$ channels} (t)
            edge                         node {$\vdots$} (t)
            edge [bend right=45]         node {} (t);
\end{tikzpicture}
\caption{Plait Network with $r$ internal nodes}
\label{fig_plait_net}
\end{center}
\end{figure}
After a simple calculation, it is not difficult to obtain $$P_e(G_2)=1-\left[\prod_{i=1}^{w}\left(1-\frac{1}{|\mF|^i}\right)\right]^{R+l}=1-(1-a)^{R+l},$$
which meets the upper bound in Theorem \ref{thm_cohe_net} with equality.
\end{rem}

However, this upper bound may require too much topological information of networks for many applications. Thus, we will give a simpler in form but looser upper bound depending on less topological information of networks.

\begin{thm}\label{thm_co_upb_1}
The failure probability of random linear network coding for the network $G$ satisfies:
$$P_e\leq 1-(1-a)^{l}(1-la)^{b}(1-ua),$$
where $\sum_{i=1}^{l}r_i=lb+u$ with $b,u$ being two nonnegative integers satisfying $0\leq u \leq l-1$ and again $r_i$ being the number of internal nodes in the chosen $w$ channel disjoint paths from $s$ to $t_i$, $i=1,2,\cdots,l$.
\end{thm}

Sometimes, we may not acquire the exact value of the sum of these $r_i$, but usually we still can obtain some topological information of networks more or less. For example, although we cannot know the sum of these $r_i$, an upper bound $n$ may be found, that is, we can find an integer $n$ satisfying $n \geq \sum_{i=1}^{l}r_i$. Let $n=l\hat{b}+\hat{u}$, where $\hat{b},\hat{u}$ are two nonnegative integers satisfying $0\leq \hat{u} \leq l-1$. Since $l\hat{b}+\hat{u} \geq lb+u$, after a simple calculation, one has
$(1-la)^{\hat{b}}(1-\hat{u}a)\leq (1-la)^{b}(1-ua).$
\begin{thm}\label{thm_upb_net}
For the network $G$, let $r_i$ be the number of internal nodes in $w$ channel disjoint paths from $s$ to $t_i$. If $\sum_{i=1}^{l}r_i\leq n$, then the failure probability of random linear network coding for the network $G$ satisfies:
$$P_e\leq 1-(1-a)^{l}(1-la)^{\hat{b}}(1-\hat{u}a),$$
where $n=l\hat{b}+\hat{u}$ with $\hat{b},\hat{u}$ being two nonnegative integers satisfying $0\leq \hat{u} \leq l-1$.
\end{thm}

In particular, for each sink node $t_i\in T$, if we can choose those $w$ channel disjoint paths which contain the minimum number of the internal nodes among the collection of all $w$ channel disjoint paths from $s$ to $t_i$, and denote this minimum number by $R_i$, then we can obtain a smaller upper bound than that in Theorem \ref{thm_co_upb_1} and having the same simple form.
\begin{cor}\label{cor_co_upb_2}
The failure probability of random linear network coding for the network $G$ satisfies:
$$P_e\leq 1-(1-a)^{l}(1-la)^{b'}(1-u'a),$$
where similarly $\sum_{i=1}^{l}R_i=lb'+u'$ with $b',u'$ being two nonnegative integers satisfying $0\leq u' \leq l-1$.
\end{cor}
\begin{rem}\label{rem_cor_co_upb}
Unfortunately, we cannot show the tightness of the upper bounds indicated in Theorems \ref{thm_co_upb_1}, \ref{thm_upb_net}, and Corollary $\mathrm{\ref{cor_co_upb_2}}$. Actually, we guess that the upper bounds are not tight. However, motivated partly by \cite{zhang-random}, we want to study the asymptotic behavior of the failure probabilities as the field size goes to infinity, because some complicated minor terms may be ignored during the derivation. So we can get a deeper understanding of the failure probability and find main factors influencing this probability. Actually, these upper bounds are asymptotically tight.
\end{rem}

Furthermore, it is apparent that the number $R$ does not exceed the number of the internal nodes $|J|$. Hence, we obtain the following theorem.
\begin{thm}\label{thm_nco_net}
The failure probability of random linear network coding for the network $G$ satisfies for $m\geq |J|$:
$$P_e\leq 1-(1-a)^{l}(1-la)^{m},$$
Particularly, if the number of the internal nodes is known,
$$P_e\leq 1-(1-a)^{|T|}(1-la)^{|J|}.$$
\end{thm}

\begin{rem}\label{rem_nco_net}
The upper bound stated in Theorem $\mathrm{\ref{thm_nco_net}}$ is also asymptotically tight.
\end{rem}

Next, we consider a linear network coding problem $\mathbf{N}^*$ which can be fully characterized by the network $G$, the source node $s$, the set $T$ of sink nodes, and the information rate $w\leq \min_{t\in T}C_t$. Thus it can be written as $\mathbf{N}^*=\{G=(V,E),s,T,w\leq \min_{t\in T}C_t\}$. Define
$$\Omega(\mathbf{N}^*)=\limsup_{\mathcal{|F|}\rightarrow \infty}\mathcal{|F|}\cdot P_{e},$$
which characterizes the limiting behavior of the failure
probability for the network as the field size goes to infinity.

Denote by $\mathcal{M}_{n,l}^*$ the set of all linear network coding problems $\mathbf{N}^*$ satisfying the following conditions:
\begin{enumerate}
  \item  the number of sink nodes is $l$,
  \item  for all sink node $t_i$, $1\leq i \leq l$, there exist $w$ channel disjoint paths from $s$ to each $t_i$ with $r_i$ internal nodes, satisfying that the sum of all $r_i$ does not exceed a fixed number $n$.
\end{enumerate}
Define
$${\Lambda_{(n,l)}^{+}}^*=\max_{\mathbf{N}^*\in \mathcal{M}_{n,l}^*}\Omega(\mathbf{N}^*),$$
which characterizes the worst case limiting behavior of the
failure probability for the network in $\mathcal{M}_{n,l}^*$.

Moreover, denote by $\mathcal{N}_{m,l}^*$ the set of all linear network coding problems $\mathbf{N}^*$ with a fixed number of internal nodes $|J|=m$ and a fixed number of sink nodes $|T|=l$. Define
$${\Omega_{(m,l)}^{+}}^*=\max_{\mathbf{N}^*\in \mathcal{N}_{m,l}^*}\Omega(\mathbf{N}^*),$$
which characterizes the worst case limiting behavior of the
failure probability for the network in $\mathcal{N}_{m,l}^*$.

From Theorem \ref{thm_upb_net} and Remark \ref{rem_cor_co_upb}, as well as Theorem \ref{thm_nco_net} and Remark \ref{rem_nco_net}, we derive the following theorem.
\begin{thm}\label{thm_limit}
For single source multicast random linear network coding, we have
$${\Lambda_{(n,l)}^{+}}^*=|T|+n\mbox{\ \ \ \ \ \ and\ \ \ \ \ \ }
  {\Omega_{(m,l)}^+}^*  =|T|(1+|J|).$$
\end{thm}


\section{Failure Probability at Sink Node}

In this section, we further give the results on the failure probability at a sink node which appeared in \cite{Guang-Fu-prop-sink-ICITIS} partly.
\begin{thm}\label{thm_cohe_sink_original}
For the network $G$ mentioned as above, the failure probability of random linear network coding at sink node $t\in T$ satisfies:
$$P_{e_t}\leq 1-\prod_{k=0}^{r}\prod_{i=1}^{w-|CUT_{t,k}^{out}|}\left(1-\frac{1}{\mathcal{|F|}^i}\right).$$
\end{thm}

This upper bound is tight for some networks such as the well-known butterfly network \cite{Guang-Fu-ButterflyNet}.
However, the upper bound may be too complicated for applications and too much topological information of the network may be required. So we give a simpler in form but looser upper bound as follows.
\begin{thm}\label{thm_cohe_sink}
For the network $G$, the failure probability of random linear network coding at sink node
$t\in T$ satisfies:
$$P_{e_t}\leq 1-\left[\prod_{i=1}^{w}\left(1-\frac{1}{|\mathcal{F}|^i}\right)\right]^{r+1},$$
where $r$ is the number of internal nodes for some collection of $w$ channel disjoint paths from $s$ to $t$.
\end{thm}

For some applications, we cannot know the number of internal nodes $r$, we can get an upper bound $n$ on the number $r$ of internal nodes, i.e., $n\geq r$. For this case, we can also analyze the failure probability at the sink node $t$.
\begin{thm}\label{thm_nco_sink}
For the network $G$, if $r \leq n$, then the failure probability of random linear network coding at the sink node $t\in T$ satisfies:
$$P_{e_t}\leq 1-\left[\prod_{i=1}^{w}\left(1-\frac{1}{|\mathcal{F}|^i}\right)\right]^{n+1}.$$
Particularly,
$$P_{e_t}\leq 1-\left[\prod_{i=1}^{w}\left(1-\frac{1}{|\mathcal{F}|^i}\right)\right]^{|J|+1}.$$
\end{thm}

The upper bounds in Theorems $\mathrm{\ref{thm_cohe_sink}}$ and $\mathrm{\ref{thm_nco_sink}}$ are also achievable for the plaint networks as the worst case.
Further, consider a linear network coding problem $\mathbf{N}^*$ and define
$$P_{e_t}^*(m,l)\triangleq \max_{\mathbf{N}^*\in {\mathcal{N}_{m,l}}^*}P_{e_t},$$
which characterizes the maximum value of the failure probability of random network coding at the sink node among all linear network coding problems $\mathbf{N}^*$ with $|T|=l$ and $|J|=m$.
\begin{thm}\label{thm_maxsink}
For linear network coding problems in ${\mathcal{N}_{m,l}}^*$,
$$P_{e_t}^*(m,l)=1-\left[\prod_{i=1}^{w}\left(1-\frac{1}{|\mathcal{F}|^i}\right)\right]^{m+1}.$$
\end{thm}


\section{Lower Bounds on The Failure Probabilities}
In addition, we can also give the lower bound on the failure probabilities.

\begin{thm}
Using random linear network coding for a single source multicast network $G$, then
\begin{itemize}
\item the failure probability at the sink node satisfies: $P_{e_t}\geq 1/|\mF|^{\delta_t+1}$,
\item the failure probability for the network $G$ satisfies: $P_{e}\geq 1/|\mF|^{\delta+1}$,
where $\delta=\min_{t\in T}\delta_t$ with $\delta_t=C_t-w$.
\end{itemize}
\end{thm}

\begin{rem}
Actually, both lower bounds above are also asymptotically achievable. Moreover, by the lower bounds on the failure probabilities, we still can obtain the conclusion proposed in \cite{Ho-etc-random}, that is, both failure probabilities tend to zero as the size of the base field goes to infinity.
\end{rem}

\section*{Acknowledgment}
The authors would like to thank Prof. Z. Zhang for his comments. This research is supported by the National Key Basic Research Program of China (973 Program Grant No. 2013CB834204), the National Natural Science Foundation of China (Nos. 61171082, 60872025, 10990011), and Fundamental Research Funds for the Central Universities of China (No. 65121007).


\end{document}